\documentclass[11pt]{amsart}
\usepackage{geometry,verbatim,cite}                
\usepackage{graphicx}
\usepackage{amssymb}
\usepackage{epstopdf}
\usepackage{graphicx}
\usepackage{amsmath}
\usepackage{mathrsfs}
\usepackage{caption}
\usepackage{subcaption}
\usepackage{color}
\usepackage{enumerate}
\usepackage{appendix}

\usepackage[normalem]{ulem}

\title[Efficient Computation of Kubo Conductivity for 2D Heterostructures]{Efficient Computation of Kubo Conductivity for Incommensurate 2D Heterostructures}

\author[D. Massatt]{Daniel Massatt}
\address{D. Massatt \\ Department of Statistics \\ University of Chicago \\ Chicago, Illionois, 60615 \\ USA.}
\email{dmassatt@uchicago.edu}

\author[S. Carr]{Stephen Carr}
\address{S. Carr \\ Department of Physics \\ Harvard University \\ Cambridge, Massachusetts 02138 \\ USA}
\email{stephencarr@g.harvard.edu}

\author[M. Luskin]{Mitchell Luskin}
\address{M. Luskin \\ School of Mathematics \\ University of Minnesota \\ Minneapolis, Minnesota, 55455 \\ USA}
\email{luskin@umn.edu}

\thanks{DM, SC, and ML were supported in
part by ARO MURI Award W911NF-14-1-0247. ML was also supported in part by
NSF grants DMS-1906129 and DMR-1922165}

\date{\today}                                           

\numberwithin{equation}{section}

\begin{document}
\maketitle

\newcommand{\ml}[1]{{\color{red} #1}}
\newcommand{\cml}[1]{{\small \it \color{red} [ML: #1]}}
\newcommand{\co}[1]{{\color{blue} #1}}
\newcommand{\cco}[1]{{\small \it \color{blue} [CO: #1]}}
\newcommand{\dm}[1]{{\color{blue} #1}} 
\newcommand{\stc}[1]{{\color{green} #1}}

\def\Xint#1{\mathchoice
{\XXint\displaystyle\textstyle{#1}}%
{\XXint\textstyle\scriptstyle{#1}}%
{\XXint\scriptstyle\scriptscriptstyle{#1}}%
{\XXint\scriptscriptstyle\scriptscriptstyle{#1}}%
\!\int}
\def\XXint#1#2#3{{\setbox0=\hbox{$#1{#2#3}{\int}$ }
\vcenter{\hbox{$#2#3$ }}\kern-.6\wd0}}
\def\mint{\Xint-}

 \newtheorem{assumption}{Assumption}
\newtheorem{remark}{Remark}
\newtheorem{prop}{Proposition}
\newtheorem{thm}{Theorem}
\newtheorem{lemma}{Lemma}
\newtheorem{definition}{Definition}
\newtheorem{corollary}{Corollary}
\numberwithin{definition}{section}
\numberwithin{thm}{section}
\numberwithin{remark}{section}
\numberwithin{prop}{section}
\numberwithin{corollary}{section}
\numberwithin{assumption}{section}
\numberwithin{lemma}{section}

\newcommand{\tbeta}{\tilde \beta}
\newcommand{\oJ}{\overline{J}}
\newcommand{\oI}{\overline{I}}
\newcommand{\oP}{\overline{P}}
\newcommand{\wH}{\widehat{H}}
\newcommand{\Id}{I}
\newcommand{\R}{\mathcal{R}}
\newcommand{\I}{\mathcal{I}}
\newcommand{\Tr}{\text{Tr}}
\newcommand{\TrN}{\text{Tr}_N}
\newcommand{\adj}{\text{adj}}
\newcommand{\per}{\text{per}}
\newcommand{\C}{\mathcal{C}}
\newcommand{\J}{\mathcal{J}}
\newcommand{\interior}{\text{int}}
\newcommand{\sch}{\mathcal{S}(\mathbb{R})}
\newcommand{\supp}{\text{supp}}
\newcommand{\Err}{\text{Err}}
\newcommand{\Imag}{\text{Im}}
\newcommand{\Real}{\text{Re}}
\newcommand{\rins}{r_{\text{ins}}}
\newcommand{\hQ}{Q}
\newcommand{\E}{\mathcal{E}}
\newcommand{\Mat}{\tilde H}
\newcommand{\G}{\mathcal{G}}
\newcommand{\Gt}{\widetilde{\mathcal{G}}}
\newcommand{\dhh}{\widehat{\delta h}}
\newcommand{\Z}{\mathbb{Z}^2}
\newcommand{\RR}{\mathbb{R}}
\newcommand{\K}{\mathcal{R}^*} 
\newcommand{\tK}{{\tilde \R^*}}
\newcommand{\Br}{B_r(0)}
\newcommand{\B}{\mathcal{B}}
\newcommand{\A}{\mathcal{A}}
\newcommand{\yt}{\tilde{y}}
\newcommand{\Rs}{\mathscr{R}}
\newcommand{\Ha}{\mathcal{H}}
\newcommand{\SN}{\mathcal{S}_N}
\newcommand{\gap}{\text{gap}}
\newcommand{\inter}{\text{inter}}
\newcommand{\intra}{\text{intra}}
\newcommand{\Gammat}{\tilde{\Gamma}}
\newcommand{\OmegaMon}{\Omega}
\newcommand{\D}{\mathcal{D}}
\newcommand{\Hmon}{H}
\newcommand{\MatSpace}{ M_{|\Omega_r|}(\mathbb{C})}
\newcommand{\M}{\mathcal{M}}
\newcommand{\Msup}{S[\widehat{H}]}
\newcommand{\nmod}{\text{mod}}
\newcommand{\hOmega}{\widehat{\Omega}}
\newcommand{\T}{\mathcal{T}}
\newcommand{\U}{\mathcal{U}}
\newcommand{\rc}{r_c}
\newcommand{\V}{\mathcal{V}}
\newcommand{\moire}{\theta}
\newcommand{\mP}{\mathcal{P}}
\newcommand{\Ll}{\mathcal{L}}
\newcommand{\bR}{\mathbf{R}^*}
\newcommand{\btR}{\mathbf{\tilde{R}^*}}
\newcommand{\interstrength}{V}
\newcommand{\X}{\mathcal{X}}
\newcommand{\Hr}{\widehat{H}_{\text{sc}}}
\newcommand{\Hstart}{\mathcal{H}_{\text{sc}}}

\begin{abstract}
Here we introduce a numerical method for computing conductivity via the Kubo Formula for incommensurate 2D bilayer heterostructures using a tight-binding framework. We begin with deriving the momentum space formulation and Kubo Formula from the real space tight-binding model using the appropriate Bloch transformation operator.
We further discuss the resulting algorithm along with its convergence rate and computation cost in terms of parameters such as relaxation time and temperature. In particular, we show that for low frequencies, low temperature, and long relaxation times conductivity can be computed very efficiently using momentum space for a wide class of materials. We then demonstrate our method by computing conductivity for twisted bilayer graphene (tBLG) for small twist angles.
\end{abstract}

\section{ Introduction }
The electronic structure of incommensurate bilayers has become a hot topic, particularly after the discovery of superconductivity in bilayer graphene with a relative twist at the so called magic angle \cite{Cao2018sc}.  {\em Twistronics}, the tuning of electronic structure by twisting stacks of 2D materials, gives a new set of parameters for tuning electronic structure, expanding the possible set of applications of these materials \cite{twistronics,Geim13}.

Incommensurate bilayers, especially for materials with small relative twist, typically require large system sizes to perform computations \cite{twistronics}. Further, given the weak van der Waals bonding between the materials, these systems are especially apt for studying via tight-binding models \cite{Castro2009}. One approach for considering such materials is through the supercell approximation \cite{terrones15}, though this can be prohibitively expensive at small angles, and leads to the computation of electronic properties for heterostructures with artificial strain since the system is not in a mechanical ground state. Real space electronic approaches have recently been developed that directly compute electronic observables such as the density of states or conductivity \cite{cances2016,massatt2017,fastkubo19}. There is also extensive literature on momentum space or $k\cdot p$ approaches, which use the monolayers' Bloch bases \cite{bistritzer2011,2019arXiv190801556C}.

In this paper, we begin with a real space tight-binding model and the real space Kubo Formula \cite{fastkubo19} and transform these using the Bloch transform into a momentum space formulation and Kubo Formula. Our approach extends the momentum space approach for the electronic density of states \cite{momentumspace17} to the formulation and computation of conductivity~\cite{momentumspace17}.  We note that a related formula is discussed in works such as \cite{Stauber_2013}. 
In this work, we are focusing on the rigorous transformation of the real space Kubo setting to the momentum space setting, and on the convergence rate of the resulting algorithm.  We note that our approach can be applied to general 2D heterostructures and is not restricted to 2D twisted heterostructures such as tBLG and can be extended to include relaxation \cite{relaxphysics18,KimRelax18,cazeaux2018energy} and trilayer systems \cite{Mora_2019}. We further demonstrate our results numerically by computing the conductivity of tBLG for several small twist angles.

The momentum space formulation directly leads to an algorithm which has far faster convergence than real space or supercell approaches for an extensive class of materials including twistronics of bilayer graphene. In addition to constructing the algorithm, we also provide a convergence estimate in terms of relaxation time and temperature. This in turn provides guidance for implementation depending on the parameters of interest.

In Section \ref{sec:rs}, we define our real space formulation. In Section \ref{sec:ms}, we derive the momentum space formulation, in Section \ref{sec:alg} we discuss the algorithm and convergence,  and in Section \ref{sec:numerics} we present simulations on tBLG to demonstrate the algorithm.

\section{Real Space}
\label{sec:rs}
Here we define the real space formulation. Each sheet is periodic in this model, so we define each sheet with respect to Bravais lattices with bases generated by the columns of $A_j$ for $j=1,2$ by
\begin{equation*}
\R_j = A_j \mathbb{Z}^2, \hspace{3mm} j \in \{1,2\}
\end{equation*}
with corresponding unit cells
\begin{equation*}
\Gamma_j = A_j[0,1)^2.
\end{equation*}
Each sheet has a finite orbital index set, $\A_j,$ that labels the orbitals associated with each lattice point in the Bravais lattice. These orbitals can be centered at any point in the unit cell thus allowing for the description of hexagon structures such as graphene and MoS$_2$ and anisotropic structures such as black phosphorous.

For $\alpha,\alpha' \in \A_j$, we define the tight-binding interaction function, $h_{\alpha\alpha'} : \R_j \rightarrow \mathbb{R}$. For $\alpha,\alpha'$ in opposite orbital index sets, we let $h_{\alpha\alpha'} : \mathbb{R}^2 \rightarrow \mathbb{R}$, which is defined over all $\mathbb{R}^2$ because of the incommensuration. We assume $h_{\alpha\alpha'}$ is smooth and exponentially localized in all its derivatives.

We then define the tight-binding degrees of freedom
\begin{equation*}
\Omega = \cup_{j=1}^2 \R_j \times \A_j
\end{equation*}
and the finite domain
\begin{equation*}
\Omega_r = \cup_{j=1}^2 (\R_j\cap B_r) \times \A_j,
\end{equation*}
where $B_r$ is the ball of radius $r$ centered at the origin.
Our tight-binding Hamiltonian operator $H$ over $\Omega$ for $\alpha \in \A_i$ and $\alpha' \in \A_k$ is given by
\begin{equation}\label{eq:conh}
[H]_{R\alpha,R'\alpha'} = h_{\alpha\alpha'}(R - R').
\end{equation}

We next construct our Kubo Formula for the real space model \cite{fastkubo19}. To begin with, we let $X_s$ be the position operator such that $(X_s)_{R\alpha} = R_s$ for $s \in \{1,2\}$, and then recall the current operator
\begin{equation}\label{eq:conxh}
[X,H]_{R\alpha,R'\alpha'} = (R-R')_s H_{R\alpha,R'\alpha'}= (R-R')_s h_{\alpha\alpha'}(R - R') .
\end{equation}
We define the current-current correlation measure $\mu_{ij}(E,E')$ \cite{fastkubo19} by the moments
\begin{equation}\label{eq:cc}
\int \phi(E)\psi(E')\,d\mu_{ij}(E,E') = \lim_{r\rightarrow \infty} \frac{1}{\#\Omega_r} \Tr_{\Omega_r} [\phi(H) \partial_i H \psi( H) \partial_j H]
\end{equation}
for all polynomials $\phi(E)$ and $\psi(E')$ (the current-current correlation measure is related
to the current-current correlation density, $\rho_{ij}(E,E'),$ by
$d\mu_{ij}(E,E')=\rho_{ij}(E,E')\,dE\,dE').$
We construct an efficient algorithm by taking moments with respect to Chebyshev polynomials $T_k(E)$ \cite{fastkubo19}.
Here $\Tr_{\Omega_r}$ means trace over $ \Omega_r \subset \Omega$ and $\#\Omega_r$ denotes the number of elements of the set $\Omega_r.$

Given the Fermi-Dirac distribution
\begin{equation*}
f_\beta(E) = \frac{1}{1+ e^{-\beta (E-E_F)}}
\end{equation*}
for $E_F$ the Fermi energy, $\beta$ the inverse temperature, and $\eta$ the inverse dissipation time, we define the conductivity function
\begin{equation}\label{eq:condfunction}
F(E,E') = \frac{i e^2}{\hbar (|\Gamma_1|+|\Gamma_2|)/2}
\frac{f_\beta(E)-f_\beta(E')}{(E-E')(E-E' + \hbar\omega + i\eta)},
\end{equation}
where $\omega$ is the frequency.
The Kubo conductivity can then be given by  \cite{fastkubo19}
\begin{equation}\label{eq:cond}
\sigma_{ij} = \int F(E,E')\, d\mu_{ij}(E,E').
\end{equation}
We can formulate the conductivity in terms of the moments \eqref{eq:cc} by expanding the conductivity function \eqref{eq:cond} in Chebyshev polynomials
\begin{equation}\label{eqn:polynomial_conductivity_approximation}
F(E,E')=    \sum_{k_1,k_2 = 0}^\infty\,\,
        c_{k_1 k_2} \, T_{k_1}(E) \, T_{k_2}(E')
\end{equation}
where $T_k(E)$ denotes the $k$th Chebyshev polynomial defined through the three-term recurrence relation
\begin{equation}\label{eqn:chebrecursion}
T_0(x) = 1, \quad
T_1(x) = x, \quad
T_{k+1}(x) = 2x \, T_k(x) - T_{k-1}(x)
.
\end{equation}

We developed a fast computational method for the conductivity \eqref{eq:cond} in \cite{fastkubo19}
by a suitably truncated Chebyshev series, which  significantly improves on the computational costs of a naive Chebyshev approximation.  We also
propose a rational approximation scheme for the low temperature regime $\eta^{-1/2} \lesssim \beta$, to remove the poles of the conductivity function \eqref{eq:condfunction}.
Chebyshev expansions will not be required in the momentum space formulation, as the Hamiltonian matrices will be far smaller than in the real space formulation, allowing for direct diagonalization.

\section{Momentum Space Formulation}
\label{sec:ms}
We next consider how to transform the real space Kubo formula to momentum space \cite{momentumspace17}. The reciprocal lattices basis vectors are generated by the columns of $2\pi A^{-T}$ giving the reciprocal lattice
\begin{equation*}
\R_j^*= 2\pi A^{-T} \mathbb{Z}^2
\end{equation*}
with corresponding unit cells (Brillouin zones)
\begin{equation*}
\Gamma_j^* = 2\pi A^{-T} [0,1)^2.
\end{equation*}
The Bloch waves for layer $1$ defined by $e^{iq_1\cdot R_1}$ for $q_1\in\Gamma_1^*$ and $R_1\in\R_1$
can be equivalently represented by
$e^{iK_2\cdot R_1}$ for $K_2\in\R_2^*$ if the heterostructure is incommensurate, and similarly for
layer $2.$
The momentum degrees of freedom space can thus be described in reciprocal space by \cite{momentumspace17}
\begin{equation*}
\Omega^* = \Omega_1^*\cup\Omega_2^* := (\R_2^* \times \A_1) \cup (\R_1^* \times \A_2).
\end{equation*}
For wave functions $\psi \in \R_j \times \A_j$, we can define the Bloch transform
\begin{equation*}
[\G_j\psi]_\alpha(q) = |\Gamma_j^*|^{-1/2}\sum_{R \in \R_j} e^{-iq\cdot R} \psi_R,
\end{equation*}
where $|\Gamma_j^*|$ denotes the area of $\Gamma_j^*.$ Likewise, we define the Bloch transform over wave functions defined on the entire heterostructure $\Omega$ by the isomorphism $\G = (\G_1,\G_2)$, where $\G_1$ and $\G_2$ act on sheet $1$ and sheet $2,$ respectively.

We now show that the momentum space operator with shift $q$ is given by \cite{momentumspace17}
\begin{equation}\label{eq:intra}
[\widehat{H}(q)]_{K\alpha,K'\alpha'} = \delta_{KK'}|{\Gamma_j^*|^{1/2}} \G_j h_{\alpha\alpha'}(q+K),\qquad
K\alpha\in \Omega_j^*,\ K'\alpha'\in \Omega_j^*\ j=1,2,
\end{equation}
for intralayer coupling and
\begin{equation}\label{eq:inter}
[\widehat{H}(q)]_{K\alpha,K'\alpha'} = \sqrt{|\Gamma_1^*|\cdot |\Gamma_2^*|} \hat{h}_{\alpha\alpha'}(q+K +K'),\qquad
K\alpha\in \Omega_1^*,\ K'\alpha'\in \Omega_2^*,
\end{equation}
for interlayer coupling where
\begin{equation*}
\hat h_{\alpha\alpha'}(\xi) = \frac{1}{(2\pi)^2}\int h_{\alpha\alpha'}(x)e^{ -ix\cdot \xi}dx.
\end{equation*}
To numerically build $\hat h$, it is most effective to build an interpolation that respects the appropriate crystal symmetry. In the case of tBLG, this should be three-fold symmetric.
The link between this momentum space operator and the real space operator is given by applying the Bloch transform:
\begin{equation}\label{eq:link}
\G [H\psi]_\alpha(q) = \left[\widehat{H}(q) \xi(q)\right]_{0\alpha},
\end{equation}
where $\xi(q)$ is the wave function 
defined by $[\xi(q)]_{K\alpha} = \G\psi_\alpha(q+K).$ See Section \ref{sec:a1} for the derivation of \eqref{eq:link}.
We define differentiation in momentum space $\partial_j$ as the derivative with respect to $q_j$, where $q = (q_1,q_2)$. In particular, we consider the operator $\partial_j \widehat{H}(q)$. This in fact is the current operator in momentum space since
\begin{equation}\label{eq:transformXH}
[\G_1 [X_j,H]\psi]_\alpha(q) = \left[\widehat{[X_j,H]}(q) \xi(q)\right]_{0\alpha}=i\left[\partial_j \widehat{H}(q) \xi(q)\right]_{0\alpha},
\end{equation}
where $[\xi(q)]_{K\alpha} = \G\psi_\alpha(q+K)$. See Section \ref{sec:a2} for the derivation.
If $A$ and $B$ are operators over $\Omega$ with the two-center form of $H$ and $[X_j,H]$
\begin{equation}\label{eq:conh}
[A]_{R\alpha,R'\alpha'} = a_{\alpha\alpha'}(R - R')\quad\text{and}\quad
[B]_{R\alpha,R'\alpha'} = b_{\alpha\alpha'}(R - R'),
\end{equation}
then $\widehat{AB}(q)=\widehat{A}(q)\widehat{B}(q)$ since, if we define $\tilde \xi(q) = \{\G[B\psi]_\alpha (q+K)\}_{K\alpha \in \Omega^*}$, we have by \eqref{eq:link} that
\begin{equation}\label{eq:ab}
\G[AB\psi](q) = \widehat{A}(q)\tilde \xi(q) = \widehat{A}(q)\widehat{B}(q)\xi(q)
\end{equation}
where $[\xi(q)]_{K\alpha} = \G\psi_\alpha(q+K).$
We showed in \cite{momentumspace17} that
\begin{equation}\label{eq:equiv}
\begin{split}
\lim_{r\rightarrow \infty} \frac{1}{\#\Omega_r} \Tr_{\Omega_r} [\phi(H)]&=\lim_{r\rightarrow \infty} \frac{1}{\#\Omega_r^*} \Tr_{\Omega_r^*} [\phi(\widehat{H}(q^*))]\\&=\nu^*\sum_{k=1}^2\sum_{\alpha\in\A_k}\int_{\Gamma_k^*}[\phi \bigl( \widehat{H}(q)\bigr)]_{0\alpha,0\alpha}dq
\end{split}
\end{equation}
for all polynomials $\phi(E)$ and $q^*\in\RR^2$, and where
\begin{equation*}
\nu^* = \frac{1}{\sum_{k=1}^2|\Gamma_k^*|\cdot |\A_k|}
\end{equation*}
and $\Omega^*_r$ is the
finite domain in momentum space
\begin{equation*}
\Omega^*_r = \Omega_{1r}^*\cup\Omega_{2r}^* :
= \left((\R_2^*\cap B_r) \times \A_1\right) \cup \left((\R_1^*\cap B_r) \times \A_2\right).
\end{equation*}
We can now apply \eqref{eq:ab} recursively to obtain that
\begin{equation}\label{eq:equiv}
\begin{split}
\int \phi(E)\psi(E')d\mu_{ij}(E,E') &=\lim_{r\rightarrow \infty} \frac{1}{\#\Omega_r} \Tr_{\Omega_r} [\phi(H) \partial_i H \psi( H) \partial_j H]\\&=\nu^*\sum_{k=1}^2\sum_{\alpha\in\A_k}\int_{\Gamma_k^*}[\phi \bigl( \widehat{H}(q)\bigr) \partial_i \widehat{H}(q)  \psi \bigl( \widehat{H}(q)\bigr) \partial_j \widehat{H}(q)]_{0\alpha,0\alpha}\,dq
\end{split}
\end{equation}
for all polynomials $\phi(E),\, \psi(E).$
We can thus equivalently reformulate the current-current correlation measure, $\mu_{ij}^*(E,E'),$ in momentum space by the moments
\begin{equation}\label{eq:ccm}
\begin{split}
\int \phi(E)\psi(E')&d\mu_{ij}^*(E,E') \\
 &= \nu^*\sum_{k=1}^2\sum_{\alpha\in\A_k}\int_{\Gamma_k^*}[\phi \bigl( \widehat{H}(q)\bigr) \partial_i \widehat{H}(q)  \psi \bigl( \widehat{H}(q)\bigr) \partial_j \widehat{H}(q)]_{0\alpha,0\alpha}dq
\end{split}
\end{equation}
for all polynomials $\phi(E)$ and $\psi(E')$ and we get
\begin{equation}
\mu_{ij} = \mu_{ij}^*.
\end{equation}
Since the Bloch transform $\G = (\G_1,\G_2)$ is an isomorphism, it follows from
\eqref{eq:link} and \eqref{eq:transformXH} that we can equivalently reformulate the conductivity
in momentum space by
\begin{equation*}
\sigma_{ij} = \int F(E,E') d\mu_{ij}^*(E,E').
\end{equation*}

\section{Algorithm}
\label{sec:alg}
In this section, we will assume the two materials have similar lattice sizes, i.e., $A_1 \approx A_2$, and we'll be interested in low temperature and large relaxation times. We also will assume frequency is low so that higher energy modes are negligible. As defined above, $\sigma_{ij}^*$ still requires the computation of a diagonal entry for an operator on an infinite-dimensional Hilbert space. To develop a computational method, we define 
the
injection operator 
by
\begin{equation*}
[P_r \xi]_{K\alpha} = \xi_{K\alpha}\delta_{K\alpha \in \Omega_r^*}.
\end{equation*}
For an operator $A$ defined over $\Omega^*,$ 
we can compute the matrix $A_r = P_r^*AP_r$.
This will be used to restrict an infinite-dimensional operator A to a finite-dimensional matrix. Indeed, we can approximate the current-current correlation measure, $\mu_{ij}^r(E,E'),$ by
\begin{equation}\label{eq:approxconductivity}
\begin{split}
\int \phi(E)\psi(E')&d\mu_{ij}^r(E,E')\\&=\nu^*\sum_{k=1}^2\sum_{\alpha\in\A_k}\int_{\Gamma_k^*}[\phi \bigl( \widehat{H}_r(q)\bigr) \partial_i \widehat{H}_r(q)  \psi \bigl( \widehat{H}_r(q)\bigr) \partial_j \widehat{H}_r(q)]_{0\alpha,0\alpha}dq
\end{split}
\end{equation}
and the approximate conductivity by
\begin{equation}
\sigma_{ij}^r = \int F(E,E')d\mu_{ij}^r(E,E').
\end{equation}

When we are interested in long relaxation times and low temperatures, the momentum space approach converges very quickly for many materials of  interest such as twisted bilayer graphene as discussed at the end of the section. Indeed, it converges so quickly that accurate results may be obtained for $r$ significantly less than the moir\'e length scale $\|A_1^{-T} - A_2^{-T}\|^{-1}$.
For example, in tBLG only wavenumbers $q$ near the Dirac points will contribute strongly to conductivity.  As a consequence of this convergence, we can reduce the domain of integration $\Gamma_k^*$ in  \eqref{eq:approxconductivity} to write a more efficient algorithm related to that used in \cite{bistritzer2011}.
In particular, our Hamiltonian can be defined over a grid of $q$-points based off the {\em incommensurate supercell}  reciprocal lattice
\begin{equation}
\R_{12}^* = 2\pi(A_1^{-T}-A_2^{-T})\mathbb{Z}^2.
\end{equation}
 This motivates us to define its unit cell of the {\em incommensurate} reciprocal moir\'e superlattice centered at $\tilde q$ to be
\begin{equation}\label{eq:restrict}
\Gamma_{12}^*(\tilde q) = \{ \tilde q + 2\pi (A_1^{-T}-A_2^{-T})\zeta  \text{ : } \zeta \in [0,1)^2\},
\end{equation}
where $\tilde q$'s will be chosen to center our regions where the integrand in \eqref{eq:approxconductivity} is significant. In the case of tBLG, we would consider two $\tilde q$'s near the Dirac points. One point would be chosen near the $K$ points for the two sheets and the other near the $K'$ points for the two sheets.

We define $(v_m,E_m)$ are the eigenpairs of $\widehat{H}_r(q)$ where $q$ is suppressed from the notation for brevity's sake. Then $\widehat{H}_r(q) = \sum_m E_m v_mv_m^*$. Then we can derive
\begin{equation}\label{eq:final}
\tilde \sigma_{ij}^r =\nu^*\sum_{\tilde q} \int_{\Gamma_{12}^*(\tilde q)}\sum_{m,m'} F(E_m,E_{m'}) \Tr[v_mv_m^*\partial_i \widehat{H}_r(q) v_{m'}v_{m'}^*\partial_j \widehat{H}_r(q)]dq.
\end{equation}
See Section \ref{sec:a3} for the derivation. 
To numerically approximate $\partial_i \widehat{H}_r(q)$, we can use any standard single variable differentiation scheme such as the centered midpoint formula to compute the derivative matrix. Note that we simply store the matrix directly, and no eigen-decomposition is used.
Finally, the integral can be uniformly discretized or stochastic sampled. Frequently to avoid bias in symmetry. stochastic sampling is preferred.

Our algorithm can achieve an exponential rate of convergence when applied to many 2D heterostructures. Firstly, we need $\|A_1^{-T} - A_2^{-T}\|$ to be small, the assumption we have used throughout this section. This obviously applies to the small twist regimes in bilayers of the same material. We further require that the Fermi energy roughly corresponds to a non-flat band regime for the monolayers. It applies well to regions with parabolic bands or Dirac points.  The technical requirements look at the collection of level sets of the monolayer band structures in terms of energy (See \cite{momentumspace17} for details).

We next consider the rate of convergence for our algorithm to compute the conductivity for such 2D heterostructures. It has been shown \cite{massatt2017} that the Green's functions of these Hamiltonians decay exponentially fast in this energy window. As a consequence, we expect that if the 2D materials and the Fermi level are as described above, we have the following rate of convergence for the our approximate conductivity to the exact Kubo conductivity:
\begin{equation}
|\tilde\sigma_{ij}^r - \sigma_{ij}| \le p(\zeta) e^{-\gamma r}
\end{equation}
where $\zeta = \max\{\beta,\eta^{-1}\},$ the decay rate $\gamma > 0$ is independent of $\varepsilon$, and $p$ is a polynomial derived from the error analysis.
The proof of this estimate follows from the same Green's function decay estimates found in Theorem 3.1 of \cite{momentumspace17}.

The outline of the algorithm is then the following:
\begin{itemize}
\item Find the required $\tilde q$'s corresponding to points near parabolic band centers or Dirac points.
\item Build $\widehat{H}_r(Q_N)$ and $\partial_i\widehat{H}_r(Q_N)$ using \eqref{eq:intra} and \eqref{eq:inter} for $\{Q_N\}$ a uniform discretization  or stochastic sampling of $\Gamma_{12}^*(\tilde q)$.
\item Compute eigenvectors and eigenvalues of $\widehat{H}_r(Q_N)$.
\item Compute the conductivity $\tilde\sigma_{ij}^r$ from \eqref{eq:final}.
\end{itemize}
We observe that this algorithm is highly  parallel in the $Q_N$ discretization and critical points $\tilde q$.

\section{Numeric example: tBLG}
\label{sec:numerics}

Applying this method to tBLG provides validation of the scheme and physically interpretable results.
As we are performing a discretization of momenta, $Q_N$, the singular nature of $F(E,E')$ even at finite $\omega$ is regularized by the size of $\eta$.
We use a $60 \times 60$ mesh sampling of $Q_N$ around each copy of the moir\'e Brillouin zone.
A value of $\eta$ corresponding to the relaxation time of graphene ($\eta \approx 10^{-6}$ eV) is too small to give smooth results in this case.
Instead, $\eta$ is taken on the order of $10^{-2}$ eV, and is a tunable parameter to ensure sufficient smoothness in the resulting $\sigma(\omega)$ curve.
For a finer mesh of $Q_N$, or if a finite element approach for interpolating between $q$ points is used, $\eta$ can be made smaller.

All results are normalized in units of the conductance of monolayer graphene, which is frequency-independent for $E_F < \hbar\omega \ll 3 eV$, and is given by $\sigma_0 = \frac{1}{4} \frac{e^2}{\hbar}$ \cite{Stauber_2008}.
As tBLG has time-reversal symmetry, only $\sigma_{xx}$ and $\sigma_{yy}$ can be non-zero, and taking into account the three-fold rotational symmetry one must have $\sigma_{xx} = \sigma_{yy}$.
As a consequence, the current-current correlation measure $d \mu$ is purely real, and $\sigma$ can be decomposed into its real and imaginary part by manipulating $F(E,E')$.
This leads to Im$(\sigma)$ not having any dependence on $f_\beta(E)$, and thus necessitating a sum over all states of the tBLG system, which removes the advantage of the continuum method.
Such a divergence can be partly corrected with a ``cancellation of infinities'' \cite{Stauber_2013}, but here we focus instead on the real part.
Thus, all results are given in normalized units of Re$(\sigma)/\sigma_0$.

\begin{figure}[ht]
\centering
\includegraphics[width=.7\linewidth]{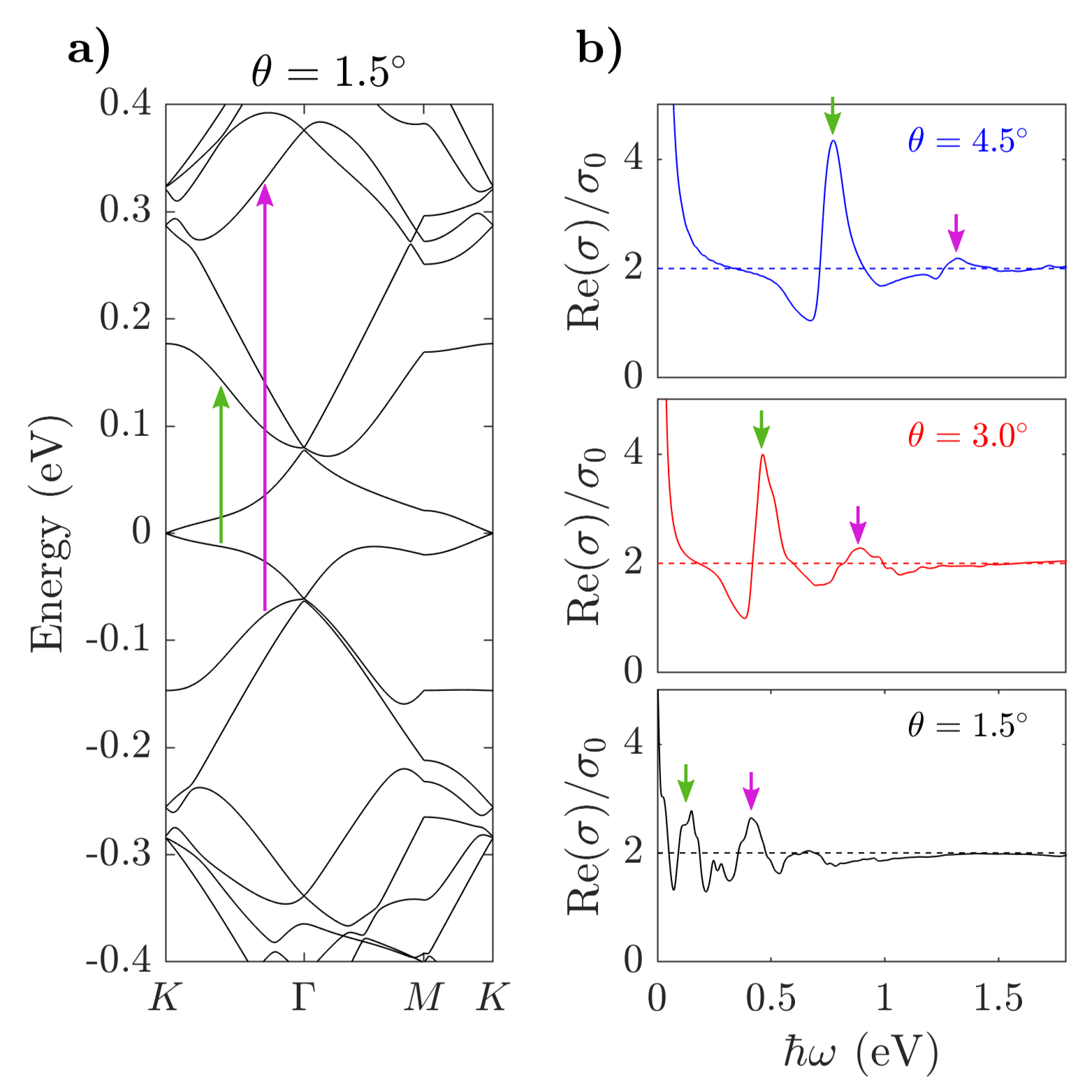}
\caption{
\textbf{a)} Band structure of tBLG for $\theta = 1.5^\circ$. The green and purple arrows highlight the interband transitions important in the conductivity calculation.
\textbf{b)} Real part of the conductivity, $Re(\sigma(0,\omega))$, normalized by $\sigma_0 = \frac{1}{4} \frac{e^2}{\hbar}$, for three different twist angles and $T = 0.3$ K.
The two inter band transitions are marked with small arrows, matching the band structure arrows.
The background value for a decoupled bilayer, $2 \sigma_0$ is shown with a dashed line.
}
\label{fig:omega_sweep}
\end{figure}

In Fig. \ref{fig:omega_sweep}b, we calculate $\sigma(E_F,\omega)$ for various $\omega$ and $E_F$ at the charge neutrality point, which is set to $0$ eV.
Evaluating $\sigma(0,\omega)$ returns reasonable results for three choices of $\theta$, with two clear peaks in the conductivity in each case.
These two peaks are associated with the interband transitions highlighted with the arrows in the band structure of Fig. \ref{fig:omega_sweep}a.
There is also a large divergence in Re$(\sigma)$ as $\omega \to 0$, which is a result of the singularity inherent in the definition of $F(E,E')$.

\begin{figure}[ht]
\centering
\includegraphics[width=.7\linewidth]{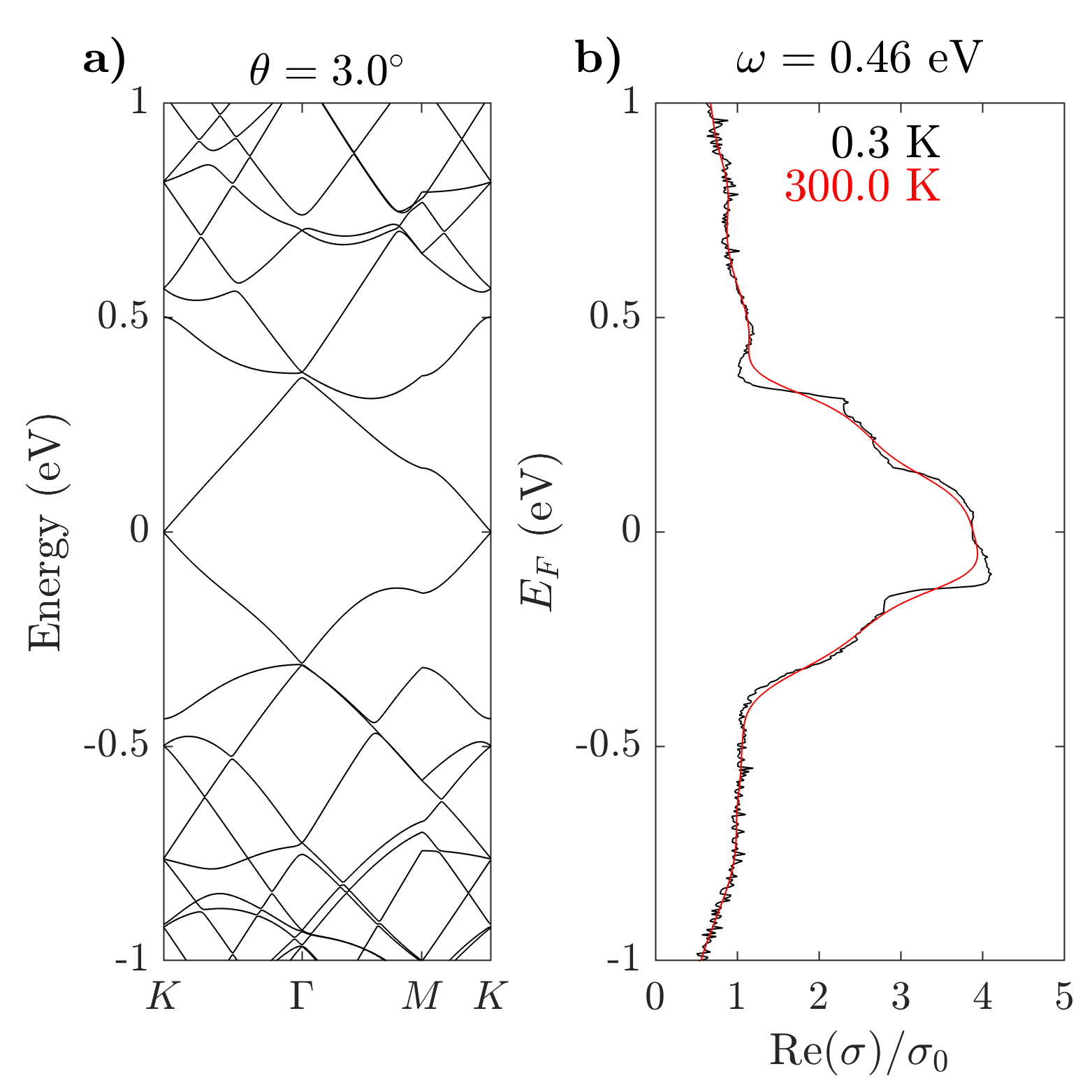}
\caption{
\textbf{a)} Band structure of tBLG for $\theta = 3.0^\circ$.
\textbf{b)} Real part of the conductivity, $Re(\sigma(E_F,\omega))$, normalized by $\sigma_0 = \frac{1}{4} \frac{e^2}{\hbar}$ as a function of the Fermi energy.
The black (red) line corresponds to T = 0.3 eV (300 eV).}
\label{fig:ef_sweep}
\end{figure}

Turning now to the dependence of $\sigma$ on $E_F$, we fix $\omega$ to the value of the first interband transition of $\theta = 3.0^\circ$ and sweep $E_F$ in Fig. \ref{fig:ef_sweep}b.
Comparing the result to the band structure at the same twist angle, it is clear that the interband tranistion is strongest at the charge neutrality point, and quickly falls off as one approaches the edges of any bands associated with that specific interband transition.
Changing the temperature from $0.3$ K to $300$ K smooths the features of Re$(\sigma)$, but otherwise has no effect.

\section{Conclusion}
In this paper, we introduced an efficient algorithm for computing conductivity in a momentum space framework and demonstrate its effectiveness for tBLG. For applicable 2D heterostructures, the algorithm converges far faster than real space approaches, and bypasses the need for supercells. We derived the momentum space model and Kubo Formula directly from the real space formulation.

The momentum space framework is very generalizable and versatile in applicability, generalizing to many incommensurate 2D systems including twisted bilayer with mechanical relaxation \cite{relaxphysics18}, and even trilayers, and providing a foundation for the development of efficient and accurate methods to compute the Kubo conductivity in 2D heterostructures.

\appendix
\section{}

\subsection{Derivation of \eqref{eq:link}.}
\label{sec:a1}
We can verify \eqref{eq:link} for intralayer coupling by setting $\alpha \in \A_1$
and observing that
\begin{equation*}
\begin{split}
[\G_1 H(\psi_1,0)^T]_\alpha(q) &= |\Gamma_1^*|^{-1/2}\sum_{R \in \R_1}e^{-iq\cdot R}\sum_{R'\alpha' \in \Omega_1} H_{R\alpha,R'\alpha'}\psi_{R'\alpha'} \\
&= |\Gamma_1^*|^{-1/2}\sum_{R \in \R_1}e^{-iq\cdot R}\sum_{R'\alpha' \in \Omega_1} h_{\alpha\alpha'}(R-R')\psi_{R'\alpha'} \\
&= |\Gamma_1^*|^{-1/2}\biggl(\sum_{R \in \R_1}e^{-iq\cdot R}h_{\alpha\alpha'}(R)\biggr)\biggl(\sum_{R' \in \R_1}e^{-iq\cdot R'}\psi_{R'\alpha'} \biggr)\\
&=  |\Gamma_1^*|^{1/2} \G_1 h_{\alpha\alpha'}(q) \G_1 \psi_{\alpha'}(q),
\end{split}
\end{equation*}
which gives \eqref{eq:link} for the intralayer coupling in momentum space \eqref{eq:intra}.

 Next we consider interlayer coupling. Letting $\alpha \in \A_1$ again, we have that
 \begin{equation*}
 \begin{split}
 [\G_1 H(0,\psi_2)^T]_\alpha(q) &=|\Gamma_1^*|^{-1/2} \sum_{R \in \R_1} e^{-iq\cdot R} \sum_{R'\alpha' \in \Omega_2}H_{R\alpha,R'\alpha'}\psi_{R'\alpha'} \\
 &= |\Gamma_1^*|^{-1/2} \sum_{R \in \R_1} e^{-iq\cdot R} \sum_{R'\alpha' \in \Omega_2}h_{\alpha\alpha'}(R-R')\psi_{R'\alpha'} \\
 &= |\Gamma_1^*|^{-1/2} \sum_{R \in \R_1} e^{-iq\cdot R} \sum_{R'\alpha' \in \Omega_2}\int\hat{h}_{\alpha\alpha'}(\zeta) e^{i \zeta \cdot (R-R')}\,d\zeta \,\psi_{R'\alpha'} \\
 &=  |\Gamma_1^*|^{1/2}|\Gamma_2^*|^{1/2}\sum_{K \in \R_1^*} \sum_{\alpha' \in \A_2}\int\hat{h}_{\alpha\alpha'}(\zeta)   \delta(\zeta - q - K) \G \psi_{\alpha'}(\zeta)\,d\zeta \\
 &= |\Gamma_1^*|^{1/2}|\Gamma_2^*|^{1/2}  \sum_{K \in \R_1^*}\hat{h}_{\alpha\alpha'}(q+K)\G_2\psi_{\alpha'}(q+K)
 \end{split}
 \end{equation*}
 by the Poisson summation formula $ \sum_{R \in \R_1} e^{i (\zeta-q )\cdot R} = |\Gamma_1^*|\sum_{K \in \R_1^*} \delta(\zeta - q - K)$
 which gives \eqref{eq:link} for the interlayer coupling in momentum space \eqref{eq:inter}.

\subsection{Derivation of \eqref{eq:transformXH}.}
\label{sec:a2}
To derive the Bloch transform of the current operator \eqref{eq:transformXH}, we first let $\alpha \in \A_1$ and consider the intralayer interaction.  We have
\begin{equation}\label{eq:xh11}
\begin{split}
[\G_1[X_j,H](\psi_1,0)^T]_\alpha(q) &= |\Gamma_1^*|^{-1/2}\sum_{R \in \R_1}e^{-iq\cdot R}\sum_{R'\alpha' \in \Omega_1} [R_j - R_j']H_{R\alpha,R'\alpha'}\psi_{R'\alpha'} \\
&= |\Gamma_1^*|^{-1/2}\sum_{R \in \R_1}e^{-iq\cdot R}\sum_{R'\alpha' \in \Omega_1} [R_j - R_j']h_{\alpha\alpha'}(R-R')\psi_{R'\alpha'} \\
&= |\Gamma_1^*|^{-1/2}\biggl(\sum_{R \in \R_1}e^{-iq\cdot R}R_jh_{\alpha\alpha'}(R)\biggr)\biggl(\sum_{R' \in \R_1}e^{-iq\cdot R'}\psi_{R'\alpha'} \biggr)\\
&= i|\Gamma_1^*|^{-1/2}\biggl(\sum_{R \in \R_1}\partial_j e^{-iq\cdot R}h_{\alpha\alpha'}(R)\biggr)\biggl(\sum_{R' \in \R_1}e^{-iq\cdot R'}\psi_{R'\alpha'} \biggr) \\
&=  i|\Gamma_1^*|^{1/2} \partial_j\G_1 h_{\alpha\alpha'}(q) \G_1 \psi_{\alpha'}(q).
\end{split}
\end{equation}

 Next, we consider interlayer coupling and let $\alpha \in \A_1$ again to derive
 \begin{equation}\label{eq:xh12}
 \begin{split}
 [\G_1[X_j,H]&(0,\psi_2)^T]_\alpha(q) =|\Gamma_1^*|^{-1/2} \sum_{R \in \R_1} e^{-iq\cdot R} \sum_{R'\alpha' \in \Omega_2}(R_j -R_j')H_{R\alpha,R'\alpha'}\psi_{R'\alpha'} \\
 &= |\Gamma_1^*|^{-1/2} \sum_{R \in \R_1} e^{-iq\cdot R} \sum_{R'\alpha' \in \Omega_2}(R_j -R_j')h_{\alpha\alpha'}(R-R')\psi_{R'\alpha'} \\
 &= |\Gamma_1^*|^{-1/2} \sum_{R \in \R_1} e^{-iq\cdot R} \sum_{R'\alpha' \in \Omega_2}(R_j -R_j')\int\hat{h}_{\alpha\alpha'}(\zeta) e^{i \zeta \cdot (R-R')}d\zeta \psi_{R'\alpha'} \\
 &= i|\Gamma_1^*|^{-1/2} \sum_{R \in \R_1} e^{-iq\cdot R} \sum_{R'\alpha' \in \Omega_2}\int\partial_j\hat{h}_{\alpha\alpha'}(\zeta)   e^{i \zeta \cdot (R-R')}d\zeta \psi_{R'\alpha'}\\
 &=  i|\Gamma_1^*|^{1/2}|\Gamma_2^*|^{1/2}\sum_{K \in \R_1^*} \sum_{\alpha' \in \A_2}\int\partial_j\hat{h}_{\alpha\alpha'}(\zeta)   \delta(\zeta - q - K) \G \psi_{\alpha'}(\zeta)d\zeta \\
 &= i|\Gamma_1^*|^{1/2}|\Gamma_2^*|^{1/2}  \sum_{K \in \R_1^*}\partial_j\hat{h}_{\alpha\alpha'}(q+K)\G_2\psi_{\alpha'}(q+K).
 \end{split}
 \end{equation}
 Putting these derivations for intralayer and interlayer coupling together gives the final result \eqref{eq:transformXH}.

  \subsection{Derivation of \eqref{eq:final}}
 \label{sec:a3}
 It is useful at this point to define in parallel to (\ref{eq:approxconductivity}) approximate and exact {\em local conductivities} in momentum space given by the local correlation measure defined by
\begin{align}
&\int \phi(E)\psi(E')d\mu_{ij,K\alpha}^r[q](E,E') = [\phi \bigl( \widehat{H}_r(q)\bigr) \partial_i \widehat{H}_r(q)  \psi \bigl( \widehat{H}_r(q)\bigr) \partial_j \widehat{H}_r(q)]_{K\alpha,K\alpha}, \\
&\int \phi(E)\psi(E')d\mu_{ij,K\alpha}[q](E,E') = [\phi \bigl( \widehat{H}(q)\bigr) \partial_i \widehat{H}(q)  \psi \bigl( \widehat{H}(q)\bigr) \partial_j \widehat{H}(q)]_{K\alpha,K\alpha}.
\end{align}
The local conductivity and its approximation are then defined by
\begin{align}
&\sigma_{ij,K\alpha}^r[q] = \int F(E,E') d\mu_{ij,K\alpha}^r[q](E,E'), \\
&\sigma_{ij,K\alpha}[q] = \int F(E,E') d\mu_{ij,K\alpha}[q](E,E'),
\end{align}
and the global conductivity and its approximation are given by
\begin{align}\label{eq:global}
&\sigma_{ij}^r = \nu^*\sum_{k=1}^2 \sum_{\alpha \in \A_k}\int_{\Gamma_k^*} \sigma_{ij,0\alpha}^r[q]dq,\\
&\sigma_{ij} = \nu^*\sum_{k=1}^2 \sum_{\alpha \in \A_k}\int_{\Gamma_k^*} \sigma_{ij,0\alpha}[q]dq.
\end{align}

We denote $P_1 = 2$ and $P_2 = 1$. For $K\alpha \in \Omega_k^*$ such that $K = 2\pi A_{P_k}^{-T}n$ where $n = (n_1,n_2)^T$ is a pair of integers, we define
\begin{equation}\label{eq:q_n}
q_n = 2\pi (A_{P_k}^{-T} - A_{k}^{-T})n.
\end{equation}
We have the identity
\begin{equation} \label{eq:sigma_shift}
\sigma_{ij,K\alpha}[q] = \sigma_{ij, 0\alpha}[q+q_n].
\end{equation}
See Section \ref{sec:a4} for the derivation of this result. 
We additionally have the approximation
\begin{equation}
\sigma_{ij,K\alpha}^r[q] \approx \sigma_{ij, 0\alpha}^r[q+q_n].
\end{equation}
An important factor in the validation of this approximation is that only energies near the Fermi energy contribute to conductivity, at least to leading order. This is because $F(E,E')$ to leading order is dominated by $E \approx E' \approx E_F$.
Consider sheet $j$ as a monolayer for a moment. Suppose $\varepsilon^j_n(q)$ is the $n^{\text{th}}$ eigenvalue corresponding to wavenumber $q$. Then it turns out only wavenumbers $q$ with corresponding eigenvalues $\varepsilon^j_n(q)$ near the Fermi energy contribute strongly to conductivity {\em in the bilayer case}.  In other words, monolayer band structure informs what wavenumbers are relevant for the bilayer system. In the case of tBLG, only wavenumbers near the Dirac cones contribute strongly when the Fermi energy is near the Dirac point. For local conductivity, this means
$\sigma_{ij,0\alpha}[q]$ becomes small if $\varepsilon^j_n(q)$ is sufficiently far from the Fermi energy for all $n$. This gives us a reduced space of wavenumbers we need to consider.

\begin{figure}[ht]
\centering
\includegraphics[width=.5\linewidth]{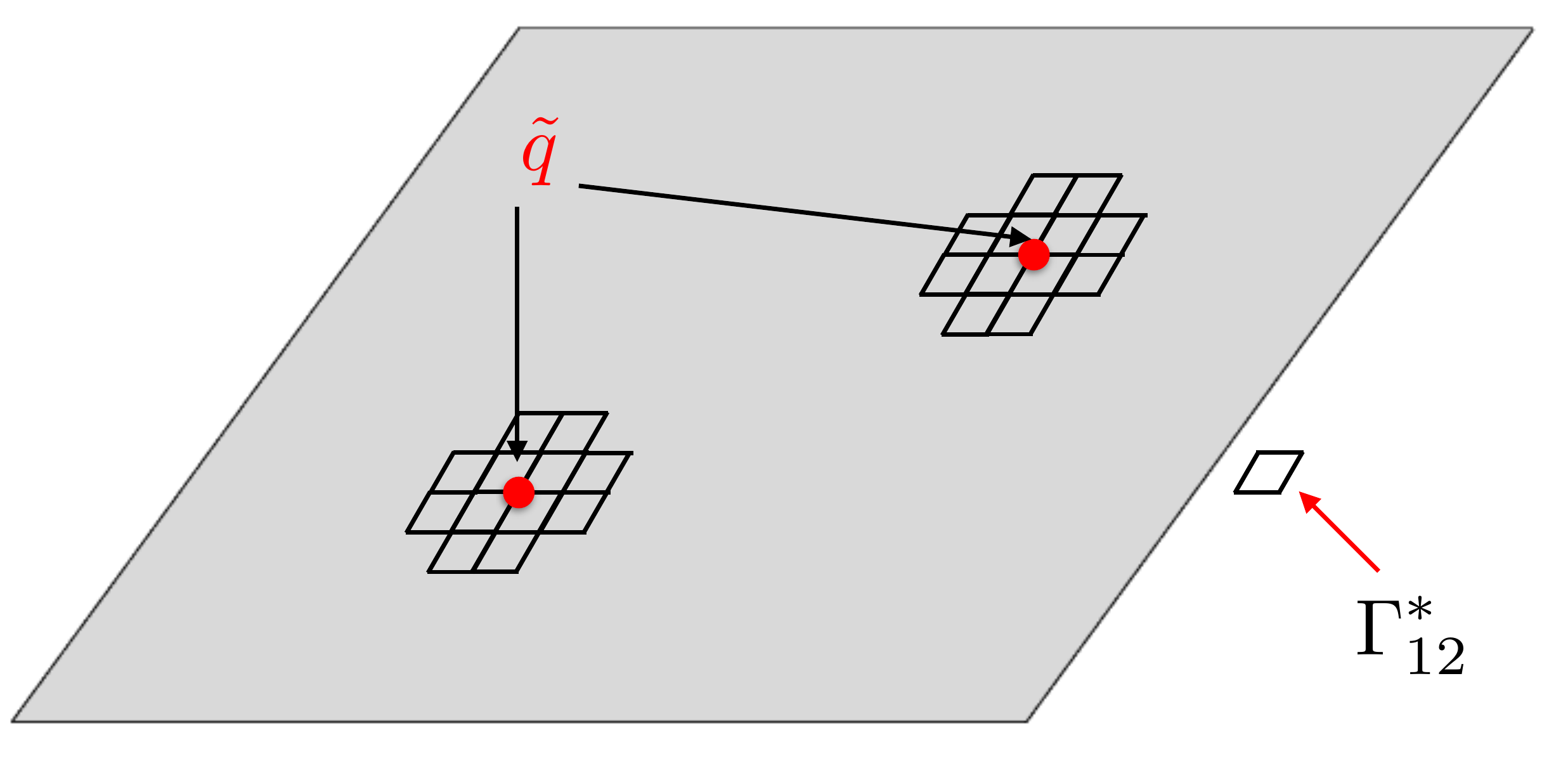}
\caption{The grey large cell is $\Gamma_1^*$. Here $\Gamma_{12}^* = 2\pi(A_1^{-T}-A_2^{-T})[0,1)^2$, a sample of the supercell reciprocal lattice unit cell. Each bottom-left vertex of the parallelograms around the $\tilde q$ points represent $\tilde q + q_n$ positions. Hence the decomposition given by \eqref{eq:set_union} breaks the regions around the $\tilde q$ points into a union of small parallelograms.}
\label{fig:supercell_BZ}
\end{figure}

As described above, the local conductivity $\sigma_{ij,0\alpha}^r[q]$ is small for $q$ far from the $\tilde q$ points.  As such, we can approximate integrals of $\sigma_{ij,0\alpha}^r[q]$ over the Brillouin zones $\Gamma_k^*$ by integrals over the much smaller isolated regions defined by the sets
\begin{equation}\label{eq:set_union}
\Gamma^*(\tilde q,k) = \bigcup_{\{n:|2\pi A_{P_k}^{-T}n| < r\}}\biggl(\Gamma_{12}^*(\tilde q) +2\pi(A_{P_k}^{-T}-A_k^{-T})n\biggr).
\end{equation}
See Figure \ref{fig:supercell_BZ}.

Using these approximations, we have
\begin{equation} \label{eq:sigma_formula}
\sigma_{ij}^r \approx  \nu^*\sum_{\tilde q}\int_{\Gamma_{12}^*(\tilde q)} \sum_{K\alpha \in \Omega^*_r}\sigma_{ij,K\alpha}^r[q]dq.
\end{equation}
See Section \ref{sec:a5} for the derivation. 
The sum in the integrand is simply a trace, which motivates us to define an approximate measure $\tilde\mu^r_{ij}$ by
\begin{equation}\label{eq:cca}
\begin{split}
\int \phi(E)\psi(E')&d\tilde\mu_{ij}^r(E,E') \\&= \nu^*\sum_{\tilde q}\int_{\Gamma_{12}^*(\tilde q)}\Tr[\phi \bigl( \widehat{H}_r(q)\bigr) \partial_i \widehat{H}_r(q)  \psi \bigl( \widehat{H}_r(q)\bigr) \partial_j \widehat{H}_r(q)]dq.
\end{split}
\end{equation}
Here we sum over the relevant regions via $\tilde q$.
For simplicity, we are assuming that the approximating integration domains are centered around points as in $\Gamma_{12}^*(\tilde q),$ though this framework can be generalized beyond such restrictions \cite{momentumspace17}.
We now have the corresponding approximate Kubo Formula:
\begin{equation}
\tilde\sigma_{ij}^r = \int F(E,E')d\tilde\mu_{ij}^r(E,E').
\end{equation}
Recall $(v_m,E_m)$ are the eigenpairs of $\widehat{H}_r(q)$ where $q$ is suppressed from the notation for brevity's sake. Then $\widehat{H}_r(q) = \sum_m E_m v_mv_m^*$. As a consequence, we have
\begin{equation}
\tilde \sigma_{ij}^r =\nu^*\sum_{\tilde q} \int_{\Gamma_{12}^*(\tilde q)}\sum_{m,m'} F(E_m,E_{m'}) \Tr[v_mv_m^*\partial_i \widehat{H}_r(q) v_{m'}v_{m'}^*\partial_j \widehat{H}_r(q)]dq.
\end{equation}

 \subsection{Derivation of \eqref{eq:sigma_shift}.}
 \label{sec:a4}
 To show this, suppose $\alpha \in \A_1$. Then let $T_K$ $(K = 2\pi A_2^{-T}n)$ be the translation of sheet 1 operator defined by
\begin{align}
&[T_K \xi]_{K'\alpha'} = \xi_{(K'-K)\alpha'} \text{ if } \alpha' \in \A_1, \\
&[T_K \xi]_{K'\alpha'} = \xi_{(K'+2\pi A_1^{-T}n)\alpha'} \text{ if } \alpha' \in \A_2.
\end{align}
Now we observe
\begin{equation}
T_K^* \widehat{H}(q)T_K = \widehat{H}(q+q_n).
\end{equation}
Note we defined the translation $T_K$ in such a way that as $n \in \mathbb{Z}^2$ varies, $\widehat{H}(q+q_n)$ varies slowly.
We next observe
\begin{equation*}
\begin{split}
 [\phi \bigl( \widehat{H}(q)\bigr) \partial_i \widehat{H}(q)  \psi& \bigl( \widehat{H}(q)\bigr) \partial_j \widehat{H}(q)]_{K\alpha,K\alpha}\\
 &=  [T_K^*\phi \bigl( \widehat{H}(q)\bigr) \partial_i \widehat{H}(q)  \psi \bigl( \widehat{H}(q)\bigr) \partial_j \widehat{H}(q)T_K]_{0\alpha,0\alpha}\\
 &=[\phi \bigl(T_K^* \widehat{H}(q)T_K\bigr) T_K^*\partial_i \widehat{H}(q) T_K \psi \bigl(T_K^* \widehat{H}(q)T_K\bigr)T_K^* \partial_j \widehat{H}(q)T_K]_{0\alpha,0\alpha} \\
 &= [\phi \bigl( \widehat{H}(q+q_n)\bigr) \partial_i \widehat{H}(q+q_n)  \psi \bigl( \widehat{H}(q+q_n)\bigr) \partial_j \widehat{H}(q+q_n)]_{0\alpha,0\alpha}.
 \end{split}
\end{equation*}
Since this holds for the local current-current correlation, it extends to local conductivity.

 \subsection{Derivation of \eqref{eq:sigma_formula}.}
 \label{sec:a5}
 We have
\begin{equation*}
\begin{split}
\sigma_{ij}^r &= \nu^*\sum_{k=1}^2 \sum_{\alpha \in \A_k} \int_{\Gamma_k^*} \sigma_{ij,0\alpha}^r[q]dq \\
&\approx \nu^*\sum_{k=1}^2 \sum_{\alpha \in \A_k} \sum_{\tilde q}\int_{\Gamma_k^*(\tilde q, k)} \sigma_{ij,0\alpha}^r[q]dq \\
&= \nu^*\sum_{k=1}^2 \sum_{\alpha \in \A_k}\sum_{K \in \R_{P_k}^*\cap B_r} \sum_{\tilde q}\int_{\Gamma_{12}^*(\tilde q)} \sigma_{ij,0\alpha}^r[q+q_n]dq \\
& \approx  \nu^*\sum_{k=1}^2 \sum_{\alpha \in \A_k}\sum_{K \in \R_{P_k}^*\cap B_r} \sum_{\tilde q}\int_{\Gamma_{12}^*(\tilde q)} \sigma_{ij,K\alpha}^r[q]dq \\
& = \nu^*\sum_{\tilde q}\int_{\Gamma_{12}^*(\tilde q)} \sum_{K\alpha \in \Omega^*_r}\sigma_{ij,K\alpha}^r[q]dq.
\end{split}
\end{equation*}

\end{document}